\documentclass[aps,prd,nofootinbib,twocolumn,superscriptaddress,floatfix,notitlepage]{revtex4-1}
\pdfoutput=1

\usepackage[normalem]{ulem}
\usepackage{graphicx}
\usepackage{amsmath,amsfonts,amssymb}
\usepackage[dvipsnames]{xcolor}
\usepackage[breaklinks,colorlinks,urlcolor=blue,citecolor=blue,linkcolor=magenta]{hyperref}
\usepackage{enumitem}
\usepackage{slashed}

\newcommand{\be}{\begin{equation}} 
\newcommand{\ee}{\end{equation}}
\definecolor{cadmiumgreen}{rgb}{0.0, 0.42, 0.24}
\newcommand{\Mpl}{M_{\mathrm{Pl}}}

\usepackage{comment}

\begin{document}
\begin{minipage}{8cm}
\vspace{-1cm}
    \begin{flushright}
IFT-UAM/CSIC-25-85
\end{flushright}
\end{minipage}

\title{Super-heated first order phase transitions}
\author{Giulio Barni}
\email{giulio.barni@ift.csic.es}
\affiliation{Instituto de F\'isica Te\'orica IFT-UAM/CSIC, Cantoblanco, E-28049, Madrid, Spain}

\author{Andrea Tesi}
\email{andrea.tesi@fi.infn.it}
\affiliation{INFN Sezione di Firenze, Via G. Sansone 1, I-50019 Sesto Fiorentino, Italy
}

\date{\today}

\begin{abstract}
    We study first order phase transitions that occur when the temperature of the system increases and we identify the conditions that lead to super-heating, a phase where the system can heat up arbitrarily. First order phase transitions with super-heating behave as inverse transitions.
     We quantify these claims by studying a prototypical example of a dark sector with a large number of interacting light bosons at finite temperature.
       Depending upon thermalisation, a super-heated phase transition in cosmology is often associated with another transition when the system is eventually cooling down, enriching the spectrum of gravitational waves from bubble collisions.
\end{abstract}

\maketitle

\section{Introduction}

The temperature of the universe has been decreasing ever since the end of inflation, when the Standard Model (SM) was reheated at its highest temperatures. Throughout the standard evolution of our Universe, the temperature has been decreasing relentlessly, possibly crossing the critical values, $T_c$, of some phase transitions. If these phase transitions were first order,
there would be a potentially detectable signal in the form of a stochastic gravitational waves (GWs) background with a characteristic peak frequency, see \cite{Caprini:2015zlo,Hindmarsh:2013xza, Hindmarsh:2015qta, Hindmarsh:2017gnf, Cutting:2018tjt, Cutting:2019zws,Gould:2021dpm, Weir:2017wfa,Hindmarsh:2020hop,Caprini:2019egz,Badger:2022nwo,Caldwell:2022qsj}.
Unfortunately, in the SM, neither the electroweak symmetry breaking nor QCD confinement are first order phase transitions. Nonetheless, these are targets of current interferometers \cite{LIGOScientific:2019vic} and PTAs \cite{NANOGrav:2023gor,EPTA:2023sfo}  as well as many future observatories such as LISA \cite{LISA} and ET \cite{Sathyaprakash:2012jk}.
This is due to the fact that it is conceivable that new physics can, for example, affect the shape of the Higgs potential, making the electroweak symmetry breaking a first order phase transition \cite{Grojean:2004xa,Delaunay:2007wb,Ellis:2018mja,Ellis:2019oqb,Bruggisser:2018mus,  Espinosa:2007qk, Beniwal:2017eik,Barger:2007im,Espinosa:2011ax,Kozaczuk:2019pet, Kurup:2017dzf,Azatov:2022tii} (see \cite{Athron:2023xlk} for a review).
More generally, other phase transitions can be present associated with the breaking of some symmetries, each with clear experimental targets at GW interferometers.

However, so far, all the dynamics have been studied in the context of transitions happening while the Universe is cooling down. This is done for good and obvious reasons, since in our Universe any trustworthy and controlled temperature rise is at most a transient phenomenon in time, that has to happen on scales necessarily shorter than a Hubble time.

As emphasised in \cite{Buen-Abad:2023hex}, however, studying
epochs where the temperature increases - while the Universe expands little - can reveal interesting new effects. Indeed, if the temperature grows above $T_c$ and then back below it, it might well be the case that a phase transition happens twice.

In this work, we explore this idea further, trying to understand what are the particle physics models that can show a phase transition when their sectors are heating up, possibly thanks to an external reservoir. In most cases we will focus on the reheating phase of the Universe, but our results will apply to more generic scenarios (such as a freeze-in type of generation).
Also, we want to identify which are those models that can be heated up maximally, in the search for large signals in GWs.

We identify scenarios of \textit{super-heating} in sectors with a large number of light bosons in thermal equilibrium, as they have the properties to generate a persistent meta-stable minimum where these bosons are actually \textit{massive}. The order parameter is a generic background field $\phi$, which, rather interestingly instead, does not need to be thermal. In these sectors, there is no \emph{a priori} upper limit on the maximal reheating temperature, but only a practical limitation induced by the {time scale for nucleation} tunnelling.\footnote{{The term “tunnelling” here is used to denote the finite-temperature decay of a metastable vacuum described by an $O(3)$-symmetric Euclidean bounce with action $S_3/T$, following the terminology originally introduced in~\cite{Linde:1981zj}.}
} The first order transition then proceeds with bubble nucleation towards the new minimum where the states are \textit{massless}. The interplay between temperature and the mass difference results in transitions of \textit{inverse} type \cite{Barni:2024lkj,Barni:2025mud}, indeed, the plasma particles gain momentum in the true vacuum where they are relativistic \cite{Buen-Abad:2023hex}. Other setups of phase transitions while heating can be found in \cite{Caprini:2011uz,Casalderrey-Solana:2022rrn,Kolesova:2023mno,Dent:2024bhi, Ai:2024cka}.

We assess the importance of two aspects underlying super-heating and heating transitions in general. First, the sector has to have a very large$-N$ number of bosons in order to populate parameter space where the effective potential for $\phi$ is under perturbative control, and the calculation of the nucleation rate is reliable. Second, the sector has to be heated-up slowly, so as to ensure that the zero temperature minimum, where the bosons are massive, becomes a metastable minimum also at high temperature. We find that efficient thermal contact with the external reservoir at temperatures below the critical value during the heating phase is essential. In this regime, the system can undergo a pair of phase transitions—an inverse transition upon heating followed by a direct transition upon cooling—and we quantify the resulting characteristic double-peaked gravitational-wave signal.

The paper is organised as follows. In section \ref{sec:superheating} we introduce the generalities of phases of superheating, putting emphasis on the form of the effective potential at finite temperature. Then, in section \ref{sec:scaleinv} we highlight a possible new physics scenario with a large$-N$ number of scalars that generates the needed scale-invariant thermal potential. We conclude the theoretical description in section \ref{sec:tree-level} by introducing mass deformation that allows us to compute the critical temperature and characterise both the reheating of the sector and its nucleation temperature. Then we discuss phenomenology in the remainder of the paper. In section \ref{sec:inverse} we describe the hydrodynamics of the transitions, and then in section \ref{sec:applications} we comment on the gravitational wave spectrum. We conclude in section \ref{sec:conclusions}.

\section{Super-heating: generalities}\label{sec:superheating}
In this section, we sketch the basic idea behind a superheating phase transition. The aim is to highlight the needed ingredients to have a potential that has a metastable minimum that persists up to extremely high temperatures. We are mostly interested in situations where the metastability of this configuration leads to a first order phase transition via {nucleation of bubbles}, as this has associated production of gravitational waves from the early universe, but we will comment on more generic endpoints.

Let us consider the example of a sector with a scalar quantum field, $\phi$, at finite temperature, $T$. As it will be evident, the sector needs additional interacting light fields, but for the moment, we do not discuss the microphysics of this setup. The typical mass scale of $\phi$ is $M$, such that at zero temperature the field has potential
\be\label{eq:zero}
V_0(\phi)=-\frac{M^2}{2}\phi^2 +\frac{\lambda_0}{4!}\phi^4\,.
\ee
For $M^2>0$ the field breaks spontaneously a $Z_2$ symmetry, since $\phi=0$ is a maximum, in a minimum $\langle \phi\rangle=f$.\footnote{Notice also that we are just interested in
classifying the properties of super-heating, the fact of using a real scalar is just for simplicity. Therefore,
we are not going to discuss the domain wall generation when $Z_2$ is spontaneously broken. Eq. \eqref{eq:zero} is
just prototypical of a sector with spontaneous symmetry breaking.}

\begin{figure}
    \centering
    \includegraphics[width=\linewidth]{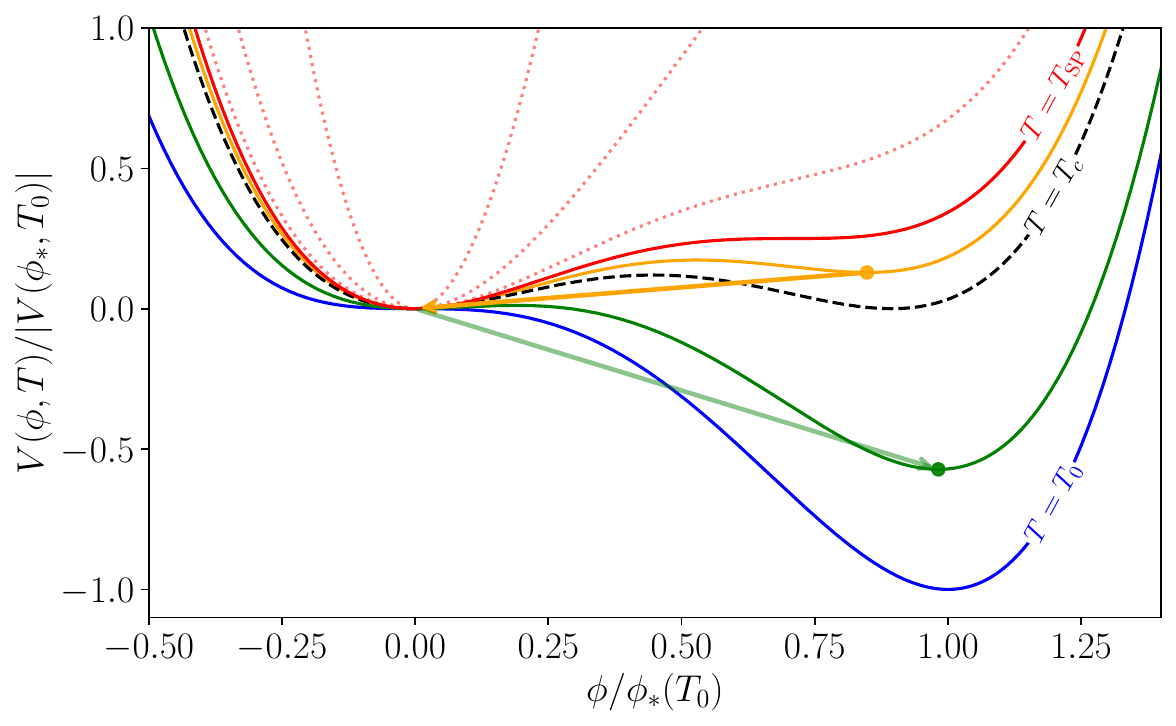}
    \caption{
Finite--temperature effective potential \(V(\phi,T)\), normalised to \(|V(\phi_\ast,T_0)|\), shown as a function of \(\phi/\phi_\ast(T_0)\) at the characteristic temperatures \(T_0\) (blue), \(T_c\) (black dashed), and \(T_{\rm SP}\) (red) (see sec. \ref{sec:tree-level}).
The plot illustrates the presence of a metastable minimum away from the origin, its degeneracy with the symmetric phase at \(T_c\), and its disappearance at the spinodal temperature \(T_{\rm SP}\).
The orange and green curves correspond to intermediate nucleation temperatures, $T_n^\leftarrow$ and $T_n^\rightarrow$ respectively (see sec. \ref{sec:nucleation}), while arrows indicate the inverse (orange) and direct (green) transition paths. Dotted light red curves denote the potential for higher temperatures in the case when superheating does not last forever.
}
    \label{fig:potential}
\end{figure}

In this work we are interested in the opposite regime of extremely high-temperature, $T\gg M$, where the effective potential for $\phi$ is subject to thermal corrections. On dimensional grounds, the possible form of the finite temperature contributions to the potential, $V_T(\phi)$, at temperatures $T\gg M$, is generically
\begin{equation}\label{eq:potentialT}
    V_T(\phi)= a_0 T^4+a_1 \phi T^3 + \frac{a_2}{2}\phi^2 T^2 +\frac{a_3}{3}\phi^3 T + \frac{a_4}{4}\phi^4\,.
\end{equation}
{The different normalizations in Eqs.~\eqref{eq:zero} and \eqref{eq:potentialT}, reflect distinct conventions for the tree-level and finite-temperature potentials, and are related by simple redefinitions of the coefficients.}
If the temperature is the largest scale, and we can neglect $M$, this is also all the potential for $\phi$, and we will discuss later in section \ref{sec:tree-level} the impact of finite $M$. In conventional field theories, we have
$a_1=0$, $a_3\leq 0$ and $a_4>0$, while the sign of $a_2$ depends on the theory \cite{Weinberg:1974hy}, but it is almost always positive. Also, usually, the field is subject to the constraint $\phi\geq 0$. \footnote{{Restricting the analysis to $\phi \ge 0$ is a matter of parametrization and does not entail any loss of generality for a quartic potential bounded from below.
}} {In fig.~\ref{fig:potential} we show the effective potential at different temperatures.
At low temperature the minimum away from the origin is the global one, while at higher temperatures the configuration at the origin becomes energetically favoured.
The superheating regime persists as long as the potential barrier separating the two minima is present at high temperatures.}

With these conditions, the presence of another minimum -- in addition to the one in the origin -- boils down to the requirements 
\begin{equation}\label{eq:barrier}
  \left\{\begin{array}{l }
       \Delta\equiv   a_3^2 - 4 a_2 a_4>0\ , \\
     a_3 < - \sqrt{\Delta}\ ,
  \end{array}\right.
\end{equation}
where the second ensures that the new minimum - which we call $\phi_*(T)$ from now on - is in the region $\phi\geq 0$. 

Notice that crucially, in this regime of ultra-high temperature, all these conditions are temperature independent. This is due to the fact that eq.~\eqref{eq:potentialT} can be rewritten in terms of a dimensionless combination {$\varphi=\phi/T$}, and the potential will just show an overall $T^4$ factor.

We are also interested in the case where $\phi=0$ is a global minimum at high temperature. This requires satisfying another constraint
\begin{equation}\label{eq:barrier-highT}
    2 a_2 a_4 >\Delta - a_3 \sqrt{\Delta}\ .
\end{equation}

Let us emphasise that if both \eqref{eq:barrier} and \eqref{eq:barrier-highT} are satisfied, then the potential \eqref{eq:potentialT} has a metastable minimum away from the origin for any value of the temperature, including extremely high values. We are therefore interested in a slightly different phase than the one mainly discussed in \cite{Buen-Abad:2023hex}.

If interested in a local minimum $\phi_{*}(T)$ at high temperature, it is convenient to consider models where $a_2$ is very large, such as to push the barrier, {located at $\phi_b(T)$}, and the local minimum away from the origin, to improve calculability. Notice that the barrier and the local minimum are always separated by $\phi_*(T)-\phi_b(T)=(\sqrt{\Delta}/a_4)\, T$, located at an average distance from the origin $\approx -a_3/(2a_4) T$ for small $\Delta$. Taking $\Delta$ as an input, we can express $a_4=(a_3^2-\Delta)/4a_2$, which leads to $a_3<-3\sqrt{\Delta}$ as an implicit new form for eq.~\eqref{eq:barrier-highT}. For this, it is convenient to define $a_3=-3 (1+\epsilon)\sqrt{\Delta}$, with $\epsilon>0$.
Therefore we have
\begin{equation}\label{eq:limits}
 \begin{split}
  V_{\rm min}&=T^4\frac{a_{2}^3 \epsilon}{\Delta  (2+3 \epsilon)^3}\ ,\\
  \frac{V_{b}}{V_{\rm min}}&=\frac{1}{4\epsilon}+O(\epsilon)\ , \quad \varphi_{b}= \frac{a_2}{2\sqrt{\Delta}}+O(\epsilon)\ ,
   \end{split}
\end{equation}
{where we recall that $\varphi_b=\phi_b/T$ and $V_b=V(\phi_b)$.}
Here $V_{\rm min}$ is equal to the gain in energy going from the metastable minimum to the origin, while $V_b$ is the value of the potential at the barrier.

\begin{figure}[t]
    \centering
    \includegraphics[width=1\linewidth]{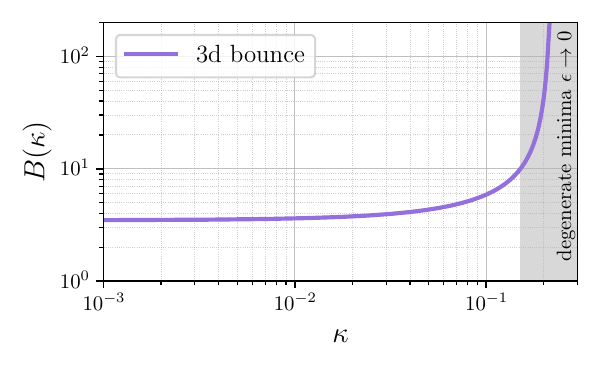}
    \caption{\label{fig:bounce3d}\it Numerical sample of $B(\kappa)$ for a 3d bounce solution, with boundary conditions as in eq.~\eqref{eq:B}.
    }
\end{figure}

Another condition to be checked is the {nucleation} rate towards the origin for a field configuration starting in the local minimum at high temperature. The action is scale invariant in the limit $T\gg M$, and reads
\begin{equation}\label{eq:S3}
 \begin{split}   
 \frac{S_3(T)}{T}&= 4\pi \frac{\tilde a_2^{3/2}}{\tilde a_3} B\left(\frac{\tilde a_2 \tilde a_4}{\tilde a_3^2}\right)\ , \\
 &=4\pi\, \frac{a_2^{3/2}}{\Delta} \frac{8 \sqrt{2}}{9   (\epsilon+2)^2(3\epsilon+2)^{3/2}}\, B\left(\frac{2(4+3\epsilon)}{9(2+\epsilon)^2}\right)\ .
 \end{split}
\end{equation}
{This expression follows from the standard result for the three–dimensional
Euclidean action of a quartic potential,
\(
S_3/T = 4\pi\, (a_2^{3/2}/a_3)\, B(a_2 a_4 / a_3^2),
\)
which is usually derived when studying {nucleation} from the origin toward a
non-zero vacuum expectation value, see \cite{Sarid:1998sn}.
In the present case, however, we are interested in the inverse process, namely
the {nucleation} from the metastable minimum at $\phi=\phi_*(T)$ back to the origin.
This is conveniently described by expanding the finite-temperature effective
potential around the local minimum $\phi_*(T)$ and identifying the coefficients
of the resulting quartic expansion.
The quantities $\tilde a_i$ denote these expansion coefficients, which can be
expressed in terms of the original parameters $a_i$ of the potential, with
$\tilde a_4 = a_4$.} Instead, here $B$ is the value of the bounce action for the dimensionless field $\varphi$ in dimensionless three-dimensional Euclidean distance $z$,
\begin{equation}\label{eq:B}
    B(\kappa)=\int z^2 dz \left[\frac{(\varphi')^2}{2}+\frac{ \varphi^2}{2}-\frac{ \varphi^3}{3}+\frac{\kappa \varphi^4}{4}\right]\ ,
\end{equation}
where $\varphi$ solves the equation of motion with boundary conditions $\varphi'(z=0)=0$ with null initial velocity and $\varphi(z\to \infty)=0$, with $B(0)=3.474$ numerically. In our example $\kappa \in [0,2/9)$, and the numerical result is shown in figure \ref{fig:bounce3d}.  We notice two interesting limits of eq.~\eqref{eq:S3}. First, when $\epsilon\to 0$ the two minima are degenerate and we are approaching the thin-wall limit, where the action diverges. In particular, we find that $B(2/9-x) \propto x^{-3/2}$ for small $x$.
Second, when $\epsilon$ is made arbitrarily large, this corresponds instead to the thick-wall limit, where $B$ approaches a constant. Curiously in this limit the action is suppressed as $\epsilon^{-7/2}$ as $\epsilon$ gets larger and larger. This suppression of the action is due to the lowering of the barrier and its vicinity to the minimum in this limit, as shown in \eqref{eq:limits}. This suppression is contrasted by possible large values of $a_2/\sqrt{\Delta}$, which push the barrier away from the origin. In both limits, and also away from them, we see that at extremely high-temperature the Euclidean action $S_3/T$ is independent of temperature. 

Therefore, for all cases where $S_3/T$ happens to be numerically large, the system -- if it finds itself in $\phi_*(T)$ -- will like to stay in the local minimum for larger and larger temperatures. We call this phase a phase of \textit{super-heating}.

\begin{figure}
    \centering
    \includegraphics[width=\linewidth]{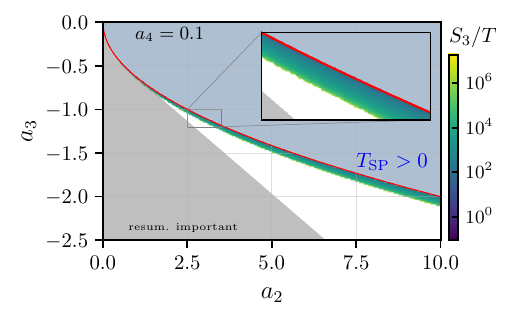}
    \caption{    \label{fig:S3_plot} \it
Bounce action \(S_{3}/T\) as a function of \(a_{2}\) and \(a_{3}\), with \(a_{4}=0.1\). Only the scale-invariant region in which the potential admits a metastable vacuum is shown. The solid red line denotes where the bounce action vanishes. The blue‑shaded area indicates the domain in which, upon adding an instability term at the origin, the spinodal temperature, eq. \eqref{eq:T spinodal}, is well defined. The grey‑shaded region corresponds to \(\phi_{b}<\pi T\), where resummation is important. }

\end{figure}

{Clearly, the thin- and thick-wall cases behave qualitatively differently with respect to super-heating (see fig.~\ref{fig:potential}), which illustrates how these two regimes correspond to markedly distinct shapes of the finite-temperature potential: in the thin-wall limit the two minima are nearly degenerate and separated by a high and narrow barrier, while in the thick-wall regime the barrier is lower and broader, bringing the metastable minimum closer to instability. 
}
In the thin-wall the action is always quite large, so the {nucleation} is rare and the system can really heat up indefinitely in the false vacuum, but the energy available for the phase transition is rather limited, being $\epsilon$-suppressed.
In the thick-wall case, the energy can be very large, but the system will like to sample smaller values of the action, leading to faster vacuum decay.

\section{Superheating from scale invariant sectors}\label{sec:scaleinv}
In the previous section we have shown the requirement on the effective potential for $\phi$. Now we identify the microphysics that can lead to this parametric behaviour.

We focus on sectors with classical scale invariance, such as a scalar or collection of scalars with only quartic interactions.
For a technical reason that will be spelt out later, we do not focus on fermions.

We believe that the simplest incarnation of super-heating is realised when we consider scale-invariant theories with $N$ scalars $S_i$, that have quartic portal couplings with $\phi$.
Very importantly, to maximize the thermal effects, no mass terms are present at the tree-level for the $S_i$, so they are massless when $\phi=0$.

In order to simplify the discussion, we assume that $S_i$ belong to the fundamental of an O($N$) symmetry group, and that such a symmetry is unbroken. The field $\phi$ is a singlet of O($N$). In this case the tree-level potential -- at zero temperature -- is given by
\begin{equation}\label{eq:potentialS}
    V_S= \frac{\lambda_{\rm mix}}{2} \phi^2 \sum_{i=1}^N S_i S_i+ \frac{\lambda_0}{4!}\phi^4 + \frac{\lambda_S}{4}\Bigl(\sum_i S_i S_i\Bigr)^2\,,
\end{equation}
where all quartics are positive. We assume that O($N$) symmetry is not spontaneously broken in such a way that when discussing the effective potential for $\phi$, we {can}
always work with $S_i=0$. In view of the large$-N$ limit that we are going to exploit, we remind that the perturbative regime of our model is attained when the combination 
\be\label{eq:thooft}
\bar\lambda\equiv\frac{\lambda_{\rm mix}\sqrt{N}}{16\pi^2}\ ,
\ee
is small. Such a condition can be derived by estimating the loop corrections to the $\phi$ quartic coupling, and the same logic can be applied to the quartic coupling for the scalars $S_i$, namely $\bar\lambda_S\equiv \lambda_S N/(16\pi^2)$. See \cite{Glioti:2018roy} for a very precise derivation and a similar setup.

When the model \eqref{eq:potentialS} is placed at finite temperature, we can compute the effective potential for $\phi$, and the coefficients $a_i$ as they appear in eq.~\eqref{eq:potentialT}. It will be sufficient to work under the assumption that only $S_i$ receive thermal corrections at temperature $T$. Therefore we just need to have thermal equilibrium in the system of the $S_i$, while we can relax this requirement for $\phi$. This is important in what follows, since $\phi$ can be treated simply as a background field. We notice that our sector can be thought of as a weakly coupled thermal bosonic CFT with a large $N$ number of states, where the coupling to the background field $\phi^2$ is the only breaking of scale invariance.

Every real bosonic degree of freedom of mass $m$ contributes to the free energy as
\be\label{eq:thermalF}
\Delta V_T(m)=\frac{T^4 J(y^2)}{2\pi^2} =\frac{T^4}{2\pi^2}\int_0^\infty dx x^2 \ln(1-e^{-\sqrt{x^2+y^2}})\ ,
\ee
with $y^2\equiv m^2/T^2$. In our case, the $S_i$ have field-dependent masses $m^2=\lambda_{\rm mix}\phi^2$, therefore eq.~\eqref{eq:thermalF} will contribute to the potential for $\phi$.  

The $J$-function is Boltzmann suppressed at low temperatures, but it is sizable at high temperatures. Let us then focus on the scenario where $T\gg m$, that is let us consider a sector with a temperature always bigger than the mass of the $S_i$. In our notation, this corresponds to expand for small $y$. Such an expansion is not analytic at $y=0$ as can be checked by noticing that the coefficient of the term $y^4$ is divergent from contributions of small momenta (see for example \cite{Laine:2016hma}). The origin of this is at the same time a feature and a worry (see section \ref{sec:resummation}). Indeed it originates from bosonic zero modes in the three-dimensional theory obtained by compactifying the time direction on a circle of imaginary radius $1/T$. In the three-dimensional theory exists an excitation of mass $m$ (potentially well below the scale $\pi T$) due to periodic boundary conditions on scalar fields, as well as a tower of heavy modes with masses multiples of $2\pi T$. 

The contribution of the zero mode results in a term linear in $T$ in the effective potential. The expression for $J$ at high-$T$ then reads,
\be\label{eq:J}
J(y^2)=-\frac{\pi^4}{45} + \frac{\pi^2}{12}y^2 -\frac{\pi}{6}(y^2)^{\frac{3}{2}} -\frac{y^4}{32}\log\Bigl(\frac{y^2}{c_B}\Bigr) +O(y^6)\,,
\ee
where $c_B=16\pi^2 \exp(3/2-2\gamma)$.
Importantly, this feature generates the non-analytic term $(y^2)^{3/2}$, which contributes to $a_3$ and has the right form to produce a thermal barrier. We heavily exploit this term to construct realistic models.

Let us notice that a contribution linear in $T$ is not present for fermions at finite temperature, since they do not have zero mode excitations in the three-dimensional theory. Both fermions and bosons contribute instead in a similar way to $a_0$ and $a_2$ (and $a_4$ with opposite signs), since these terms do not depend much on the contributions from zero modes.

All in all, by expanding at high-temperature the thermal functions for the $S_i$, we get the following contributions to the $\phi$ potential of eq.~\eqref{eq:potentialT}
\begin{equation}\label{eq:matching}
  \begin{split}
 a_0&=-\frac{\pi^2}{90}N,\quad  a_1 =0\,,\quad a_2=N\frac{\lambda_{\rm mix}}{12},\\
 a_3&= - N \frac{\lambda_{\rm mix}^{3/2}}{4\pi},\quad a_4 =\frac{\lambda_0}{6}-N\frac{\lambda_{\rm mix}^2}{16\pi^2}\ell\,,      
  \end{split} 
\end{equation}
where $\ell\equiv \log(\lambda_{\rm mix}\phi^2/(T^2 c_B))$\footnote{Notice, that adding the Coleman-Weinberg potential for $\phi$ will lead to a $\phi$-independent logarithm.}. 

Let us comment on the size of these coefficients at large$-N$. Clearly, the thermal pressure of the particles is enhanced by $N$. By contrast, the other coefficients show a different behaviour with $N$ when expressed in terms of $\bar\lambda$, which we hold fixed and perturbative. In particular, we have that $a_2\propto \bar\lambda\sqrt{N}$ and $a_3\propto 1/\sqrt{N}$. This shows that we can have parametrically large values of $a_2$ while still being perturbative. Also, since given the scaling of $a_3$ we cannot really take the limit $N\to \infty$, as this imposes an upper bound.

In the model under consideration, the conditions in Eqs.~\eqref{eq:barrier} and \eqref{eq:barrier-highT} are independent of $N$, yet together they imply the constraint $2\sqrt{2\lambda_0}/3<16\pi\bar\lambda<\sqrt{\lambda_0}$. At the same time, requiring the barrier lie sufficiently far from the origin—so that re-summation can be avoided (cf.\ the next subsection)—yields $16\pi^2\bar\lambda\lesssim\sqrt{N}/9$. Furthermore, since the bounce action scales as $S_3/T\propto N^{1/4}$, there is an upper bound on $N$ to ensure the {nucleation} condition is satisfied. This, as exemplified by figure \ref{fig:S3_plot}, leads to $\bar\lambda\approx 0.015$ and $N\approx 250$.

Notice that the quartic coupling $a_4$ in the effective potential already receives a contribution from the tree-level coupling $\lambda_0$, not affected by large$-N$ constraints. We notice that the regime where $\lambda_0$ controls $a_4$ is given by the condition $\lambda_0 \approx 6 a_4 \gtrsim 16\pi^2 \bar\lambda^2 \approx 0.08 a_3^2/a_2$. Such a condition applies in the region of interest of figure \ref{fig:S3_plot}.

This discussion has a very strong analogy with the one of Ref.~ \cite{Glioti:2018roy}, where the authors have identified a mechanism to break electroweak symmetry at arbitrarily high temperatures. In analogy, here we still want a symmetry-breaking minimum at high temperatures, but differently from that work, we are after a metastable minimum leading to a first order phase transition. This leads us to work in a limit where resummation is not important as not to suppress the contribution to the barrier as we discuss next.

\subsection{High-temperature and zero modes}\label{sec:resummation}
{We treat $\phi$ as a background field and not as a fully thermalized degree of freedom. 
This choice allows us to isolate in a controlled way the thermal origin of the barrier responsible for the super-heating mechanism. 
In our framework, the barrier arises from the non-analytic high-temperature contribution of the thermalized spectator bosons, which generates an effective cubic term $\sim \phi^3 T$ in the finite-temperature potential. 

For this reason, we consider $\phi$ as weakly coupled to a thermalized sector, acting as an order parameter probing the thermal structure generated by the bath fields. 
This assumption is not merely a simplifying choice, but is essential to cleanly identify and characterise the super-heating mechanism studied in this work.
}

{In this regard,} the only way to have a metastable minimum at high temperature, via renormalizable interactions, is to exploit the term linear in $T$ in the effective potential for $\phi$, balanced against the term proportional to $T^2$. \footnote{Allowing for non-renormalizable interactions of dimension $d_\Lambda$, we can have other terms of size $T^4(T/\Lambda)^{d_\Lambda}$, where $\Lambda$ is the effective scale.} 
For us, it is then very important to assess the robustness of the matching condition in eq.~\eqref{eq:matching} for $a_3$. And this leads us to the worry introduced by the presence of bosonic zero modes. As is well known, the zero modes can enter a strongly coupled regime even if their couplings in the four dimensional theory are perturbative, since the small parameter of the perturbative series at finite temperature is $\varepsilon\approx \pi \bar\lambda_S T/m$, using the $S_i$ self-quartic for the estimate. For arbitrarily small values of $m$, which in our case corresponds to background field value $\phi$ close to the origin, the perturbative expansion is clearly subject to large corrections, which likely need to be re-summed to all orders, if at all possible. In usual situations, the procedure of re-summation allows eventually for a well-behaved limit $m\to 0$ for the zero mode, for example by shifting it with the 1-loop thermal self-energy \cite{Arnold:1992rz,Parwani:1991gq} (see also \cite{Curtin:2016urg,Bittar:2025lcr} for more recent updates).

This just explains our worry: if we are in a situation where $\varepsilon$ is not tiny, the size of $a_3$ is subject to a great reduction, which we want to avoid. 

In practice, we want to simply avoid the need for re-summation, that is, we always work in the limit $\varepsilon\ll 1$. At our benefit, let us recall that $\phi$ is just a background field, so for us it is enough to explore field value of $\phi$ such that $\varepsilon$ is small, while still expanding the $J$-function for $T\gg m$. Then, it will be important to consistently check that we are in a situation of thick-wall phase transition where we do not need to explore the region close to the origin, or not even exactly know the potential there, but rather be in a limit where $\phi\gg T$. See also \cite{Buen-Abad:2023hex} for similar considerations. Fortunately, this condition is achieved with a large $a_2$, which has at least the virtue of growing with $N$.

\section{Superheating and critical temperature}\label{sec:tree-level}
In the sections \ref{sec:superheating} and \ref{sec:scaleinv} we have outlined the two main ideas behind super-heating, that is, the need for a scale invariant (conformal) sector with a scalar $\phi$ with a small quartic and a large$-N$ number of light bosons. If these conditions are met, super-heating can occur at high-temperature if the system find itself in the minimum $\phi_*(T)$. Concretely, we imagine a scenario where the large$-N$ sector establish thermal equilibrium during a phase where an external reservoire act as a source of energy to increase the temperature from zero to a maximal value $T_{\rm max}$. When the system reaches $T_{\rm max}$, then the temperature decreases until it reaches zero again. We want to understand what is happening during the temperature raise, as well as during the inevitable cooling. 

In particular, the main point here is to explain how the system can find itself in the minimum away from the origin at $\phi_*(T)$. Interestingly this can be done easily by bringing back into the game the mass instability \eqref{eq:zero} that we have discarded at high temperatures.

Indeed, there will be a range of temperatures, during the heating, where we cannot at all neglect the instability of eq.~\eqref{eq:zero}. Therefore, we are led to see what are the implications of the mass instability, as this is strongly connected with the existence of a critical temperature, where two degenerate minima appear. Such a mass term renders all scenarios realistic and feasible.

Practically, the presence of the instability, can be traced back to all our formulas simply shifting $a_2\to a_2(T)= a_2 - M^2/T^2$. And notice that the conditions of eq.s~\eqref{eq:barrier} and \eqref{eq:barrier-highT} can still apply with $a_2(T)$ in place of $a_2$.

However, the instability is a major deformation of the potential and of the dynamics, as it introduces a temperature $T_0$ defined by $a_2(T_0)=0$, which identifies the alleged second-order phase transition end point, $T_0=M/\sqrt{a_2}$, {see blue curve in fig.~\ref{fig:potential}}.

At the same time, now it is also possible to identify a critical temperature $T_c$, where the minimum in the origin and the one in $\phi_*(T_c)$ are degenerate, {see black dashed curve in fig.~\ref{fig:potential}.} The condition is $a_2(T_c)=\frac{2}{9} a_3^2/a_4$, which reads off as 
\begin{equation}
    T_c = \frac{M}{\sqrt{a_2}} \left(1-\frac{2}{9}\frac{a_3^2}{a_2 a_4}\right)^{-1/2} > T_0\,.
\end{equation}
For temperatures above $T_c$ -- if $a_2(T)$, $a_3$ and $a_4$ satisfy the constraints -- the minimum in $\phi_*(T)$ is potentially unstable towards decay into the origin of the potential.

It is interesting now to follow the time (or temperature) dependence of the action, since now there is a scale in the system. For all temperatures around $T\gtrsim T_c$, we expect a modification from the second line of eq.~\eqref{eq:S3}, induced by $a_2(T)$. However, for $T\gg T_c$ we should recover the scaling limit. Notice, importantly, that the conditions for the existence of the second minimum above the origin are now $T-$dependent and can be modified. In particular we can trade $M$ for $T_c$ with the above relation. For $T$ just marginally bigger than $T_c$, we have that $S_3/T$ is very large (exactly infinite at the critical temperature), eventually at high-$T$ it will reach the expression of \eqref{eq:S3}. The question here is if there is a minimum for the action at finite $T$, or if the minimum is attained asymptotically at very high-temperatures. In the latter case, the most favourable condition for {nucleation} is at high temperature.

\subsection{Heating in the meta-stable minimum}
At zero temperature,  the system finds itself in the minimum, which, with the previous conventions, is located at $\langle \phi\rangle\equiv f=M/\sqrt{a_4}$. In this field configuration, the particles all have masses proportional to $f$ times a weak coupling. The scale $f$ can be used to measure the temperature $T_0$ as well as the critical temperature $T_c$. 
On the other side, the masses of the particles in the $T=0$ vacuum are $M_\phi=\sqrt{2a_4}f$ and $M_{S_i}=\sqrt{\lambda_{\rm mix}}f$.

We believe that $T_0$ is now an important parameter, since for all temperatures $T<T_0$ there is only one minimum away from the origin. In this way, if the sector establishes thermal equilibrium below $T_0$ we are sure that, when heated up further, it will explore the phase of super-heating evolving from $f$ to $\phi_*(T)$, never leaving the minimum until the action drops below the nucleation threshold. This way, we are sure to have the system in the metastable minimum at high temperature. We want, therefore, a smooth evolution of $\phi$ as
\be
f \to \phi_*(T)\ , \quad\quad \text{during heating}.
\ee
By construction, we see that $T_0$ is always smaller than $M_\phi$, which suggest that thermal corrections due to a thermal population of $\phi$ are Boltzmann suppressed. Indeed we did not include them at all in the effective potential for $\phi$ in eqs.~\eqref{eq:potentialT} and \eqref{eq:matching}. {As discussed, it is the spectator fields $S_i$ that thermalise}
in the background of $\phi$. For this reason, it is just important to check the condition
\be
T_0\gtrsim M_S =\sqrt{\lambda_{\rm mix}} f\ ,
\ee
which guarantees that the effective potential of eq.~\eqref{eq:potentialT} and \eqref{eq:matching} is valid already before the system reaches the temperature $T_0$. This requirement enforces a relation between the critical temperature and $\lambda_{\rm mix}$, leading to
\be
16\pi^2\bar\lambda \lesssim {\sqrt{2\lambda_0}} \ .
\ee
We note that this is only marginally violated in our parameter space, implying that $M_S \sim T_0$ is already sufficient to guarantee that our treatment remains qualitatively valid. In this regime, a thermal population of the $S_i$ can still form around $T_0$. While this could slightly modify the dynamics near $T_0$, we expect our high-temperature expansion to remain a good approximation in the thick-wall regime. For the thin-wall case, where $T_c/T_0 \simeq 1$, the situation might require a dedicated treatment. We regard this as a mild tension in the model and leave a more refined analysis for future work.

\subsection{Spinodal temperature}
So far, we have outlined super-heating in the case where the existence of the second minimum at high temperature is set by conditions that hold for arbitrarily high temperatures. This leads to maximal super-heating up to the available $T_{\rm max}$, whose magnitude depends on the reheating process (see next section). We notice, however, that for some choice of parameters, super-heating can last indefinitely. This means that the system - after the temperature eventually drops below $T_{\rm max}$ - will cool down never restoring the symmetry at $\phi=0$.

To avoid this, we can relax the conditions for the existence of the barrier and minima of \eqref{eq:barrier} and \eqref{eq:barrier-highT}. In particular while holding on \eqref{eq:barrier-highT} with $a_2(T)$, we can just require $\Delta$ just to be positive up to some temperature $T_{\rm SP}$ where it becomes zero. At this point, the minimum $\phi_*(T_{\rm SP})$ can be a saddle point or it can just disappear. This brings to the condition $a_2(T_{\rm SP})=\frac{1}{4} a_3^2/a_4$, which reads off as 
\begin{equation}
\label{eq:T spinodal}
    T_{\rm SP} =  \frac{M}{\sqrt{a_2}} \left(1-\frac{1}{4}\frac{a_3^2}{a_2 a_4}\right)^{-1/2} > T_c\,.
\end{equation}
We notice that it is defined as soon as $\Delta<0$, i.e. $a_3^2<4a_2a_4$. {Therefore, at the spinodal temperature, the barrier is disappearing, and the minimum away from the origin ceases to exist, as one can see from the red solid curve in fig.~\ref{fig:potential}.}
\begin{figure*}[t!]
    \centering
    \includegraphics[width=1\textwidth]{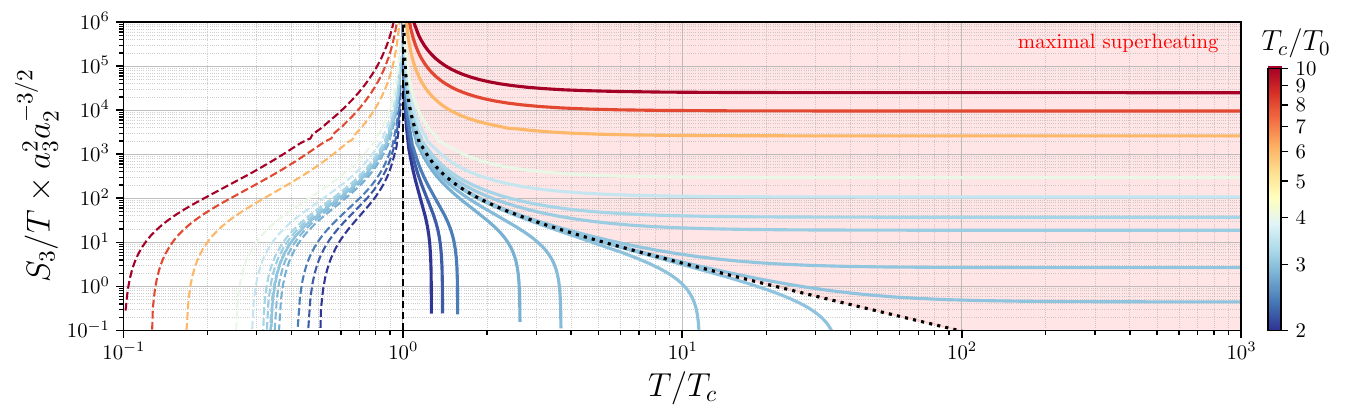}
    \caption{\it 
Bounce actions for heating (solid) and cooling (dashed) phase transitions, colour-coded by critical temperature $T_c/T_0$.  Each solid curve at a given colour corresponds to its matching dashed curve upon cooling. Solid lines stop at the spinodal temperature, eq. \eqref{eq:T spinodal}, when it exists, while dashed ones stop at $T_0$.  The colorbar shows $T_c/T_0$. We highlight that the red‐shaded region ($T_c/T_0>3$), above the black dotted line, marks the maximal superheating phase where the barrier persists to arbitrarily high $T$ and it is disjoined from the rest of the parameter space. In this phase, the action, at large temperature, converges to the scale‐invariant value.}
    \label{fig:S3_temp}
\end{figure*}

\subsection{Models for the reheating sector}\label{sec:reheating}
Let us finally comment on how the temperature can rise, since the decrease is easy to obtain thanks to the expansion of the Universe. We imagine that the sector described by $\phi$ and $S_i$ is populated by extracting energy from a ``reheating sector", $R$, with large energy density that is passed onto our sector with a rate $\Gamma_R$. At the same time, in the reheated sector, there is rate of thermalization, $\Gamma_T$, which has to be the fastest scale. Such a rate converts into thermal energy, at a temperature $T$, all the bits of energy extracted from the reservoir $R$. Our discussion here adheres to the one in \cite{Buen-Abad:2023hex}. 

Let us just emphasise a few points that we believe are new and relevant to our construction. 

It is important to even talk about the super-heating phase, that the system of $S_i$ thermalises below $T_0$, and not that it is immediately burnt at $T_{\rm max}$, otherwise we cannot argue in favour of a smooth growth from $f$ to $\phi_*(T)$. For this reason we want to focus on a scenario where $\Gamma_R\ll H$, such that the energy density of $R$ is transferred up to a time interval equal to $\Gamma_R^{-1}$. This way, the energy density of $S_i$ is growing in time up to when all $R$ has decayed. This will have an impact on the gravitational wave signal (see section \ref{sec:applications}).

Another aspect worth mentioning is that thermalisation is just controlled by $\lambda_S$, which needs to be smaller than $16\pi^2/N$. This is another reason why, realistically, we cannot take the $N\to \infty$ limit.

If all these conditions are met, on time scales smaller than $1/\Gamma_R$, the energy density of $S_i$ grows linearly with time $t\ll 1/\Gamma_R$, and the temperature raises as $T(t)\propto (t\Gamma_R)^{1/4}\sqrt{H_* \Mpl}/N^{\frac{1}{4}}$, where $\Mpl$ is the Planck scale and $H_*$ is the Hubble scale at the onset of the heating phase. 
There is a short window where the temperature is rising, which we exploit next. We can think of this as the reheating phase of our Universe mentioned in the Introduction. This way $H_*$ would be the typical value of the Hubble scale after inflation during a matter dominated era, while the inflaton or any other reheating field is dominating the energy density prior to decay into our sector (and the SM).

\subsection{Low-temperature phase transition below $T_c$}
For all cases where nucleation happened before the system cools below $T_{c}$, the system will find itself in the origin while the temperature is decreasing adiabatically. At this stage, we can then ask whether the system will {transition} to the minimum away from the origin when the temperature drops back below $T_c$. Notice, that here we face an even more worrying aspect of the dynamics of the zero modes already discussed in section \ref{sec:resummation}. Indeed, close to $T_0$, the effective mass of the zero mode is extremely small compared to the temperature, making unreliable all computations that need to know precisely the effective potential at the origin. 

Therefore, in section \ref{sec:nucleation}, we refrain from deriving strong conclusions in this regime, except that in the case where there is a clear separation $T_c\gg T_0$ and the {nucleation} proceeds in the thick-wall limit, where the details of the potential are irrelevant. In all cases where nucleation happens well before $T_0$ we can argue in favour of another first order phase transition.

In Fig.~\ref{fig:S3_temp} we compute the bounce actions for heating and cooling transitions, above and below $T_c$ respectively, showing that each heating‐induced transition (solid lines) is in principle mirrored by its counterpart on cooling (dashed lines). As the figure
illustrates, for the heated phase transitions, the parameter space is divided into two regimes separated by the limiting value \(T_c/T_0=3\) found when $\Delta=0$ (dotted black line).  When \(T_c/T_0>3\), the barrier survives at arbitrarily high temperature (maximal superheating branch in shaded red); for \(1<T_c/T_0<3\), a finite spinodal \(T_{\rm SP}\) appears and the barrier vanishes, allowing the transition to complete just below it. 

\subsection{Nucleation temperature}\label{sec:nucleation}
\begin{figure*}[t]
    \centering
    \includegraphics[width=0.49\linewidth]{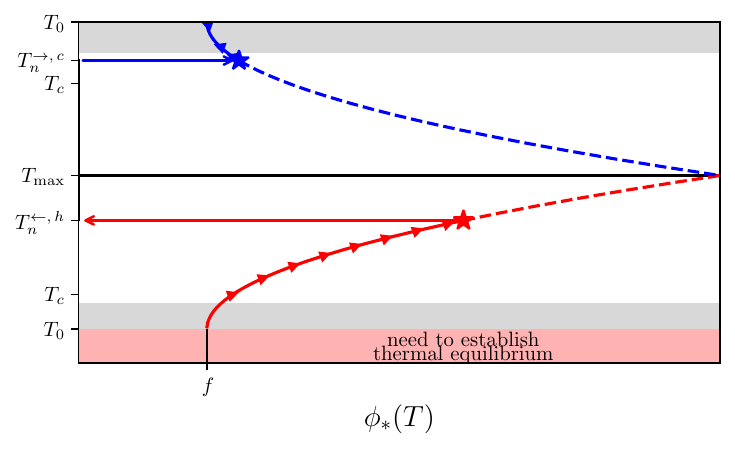}\includegraphics[width=0.49\linewidth]{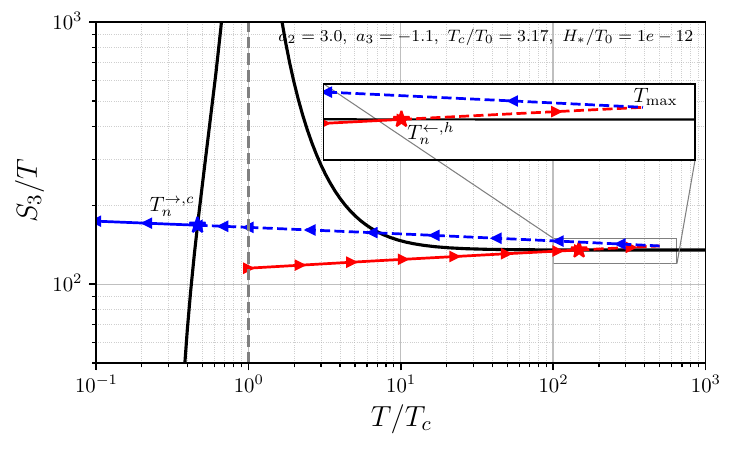}\\~~\includegraphics[width=0.49\linewidth]{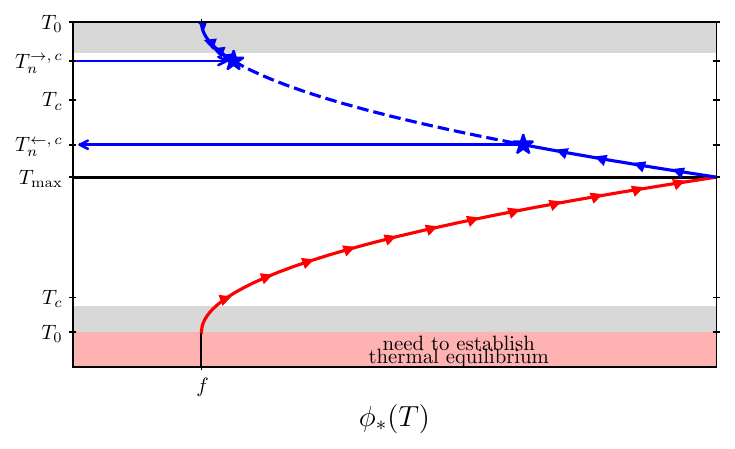}\includegraphics[width=0.49\linewidth]{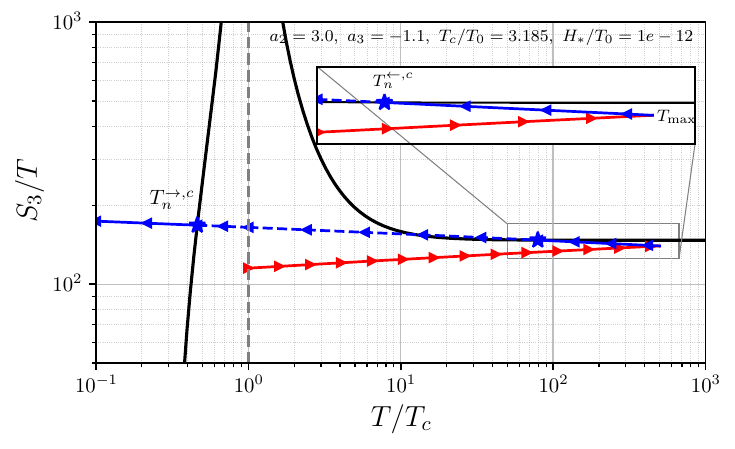}
    \caption{\it
Illustration of the model with two phase transitions: while heating (top row) and cooling (bottom row). Heating is represented in red, while cooling is in blue.
\textbf{Left column:} Schematic trajectories of the order‐parameter value. Solid curves show the actual evolution, arrows on the lines indicate the direction of time, and stars mark the nucleation temperatures for the inverse ($\leftarrow$) and direct ($\rightarrow$) transitions. Dashed extensions depict the hypothetical path if {nucleation} did not occur. The red shaded area at the bottom indicates the thermalisation of the system below $T_0$.
\textbf{Right column:} Corresponding bounce action \(S_3/T\) (black curve) with the nucleation criterion from Eq.~\eqref{eq:nucleation} overlaid (coloured lines for different Hubble rates). Solid segments represent the realised evolution.
In both heating and cooling cycles, once the inverse transition occurs, the direct transition inevitably follows.%
}
    \label{fig:branch}
\end{figure*}

Nucleation can occur for a variety of directions. We refer to $\leftrightarrows$ for transition towards the origin and away from it, respectively. They can happen both when the system is cooling ($c$) or heating ($h$).

For all practical conditions, the Hubble scale is constant while the system is heating up. Therefore, the nucleation condition, during heating, is given by
\begin{equation}
\label{eq:nucleation}
    T^4 \exp(-S_3/T)\approx H_*^4\, , \quad \text{(heating)}
\end{equation}
while during cooling $H_*$ is replaced by $H(t)$.

Let us first discuss the heating phase. Notice that since $T$ grows in this phase, we can identify a solution where $T_n^{\leftarrow,h}$ is smaller than $T_{\rm max}$, by comparing $S_3/T$ to $4\ln (T/H_*)$. Let us notice, that if there is no solution for $T_n^{\leftarrow,h}$, then the system will stay in that minimum while the temperature is cooling.

If a $T_n^{\leftarrow,h}$ exists, the system can even {transition} back to $\langle \phi \rangle \approx f$ when, after $T_{\rm max}$, the temperature cools adiabatically with the expansion of the universe. In that case, assuming radiation dominance, the Hubble scale is controlled by the SM temperature, which can differ by a factor of $1/\xi$ compared to $T$, usually $\xi\ll 1$. In that case, $S_3/T$ has to be compared with $\approx 4\log \Mpl/T$, and there might be a case where $T_n^{\rightarrow,c}$ exists well before $T_0$. We require $T_n^{\rightarrow,c}\gg T_0$ to make reliable estimates based on the shape of the potential just away from the origin.  It is amusing to see that there are either zero or two phase transitions, but not just one. Often, the two are also both strongly first order. We show in figure \ref{fig:branch}-left a region of parameter space where both $T_n^{\leftarrow,h}$ and $T^{\rightarrow,c}$ exist.

Else, if a $T_n^{\leftarrow,h}$ does not exist, most likely we have ended up in a region of maximal superheating where $S_3/T$ approaches a constant (see figure \ref{fig:S3_plot}). In this case, the action is just too big, since $4\log T/H_R$ just falls short of activating nucleation before $T_{\rm max}$.  We notice, and this was not present in \cite{Buen-Abad:2023hex}, that when the temperature drops below $T_{\rm max}$, the condition for nucleation compares $S_3/T$ almost constant and $4\ln (T/H)$, which grows as temperature drops. This implies the possible existence of a solution for a new nucleation can happen with a $T_n^{\leftarrow,c}$ while cooling. Eventually there will be another transition $T_n^{\rightarrow,c}$ to the minimum in $f$ while cooling. Again, there are either zero or two phase transitions. We show this parameter space in figure \ref{fig:branch}-right.

As a final comment, we notice that in the region of parameter space where a finite spinodal temperature \(T_{\rm SP}\) exists, the inverse transition during heating is guaranteed to occur, and the corresponding direct transition on cooling inevitably follows. However, because the direct cooling transition nucleates at temperatures close to \(T_0\), the potential barrier lies too much near the origin where perturbation theory is not under control; consequently, at sufficiently low temperatures the exact shape of the potential near \(\phi=0\) is uncertain, and one cannot definitively determine the order of this second transition.

\section{Superheating: inverse transition}\label{sec:inverse}
The phase transition at high temperature brings the system to a vacuum with a large$-N$ number of thermal states. Since they are massless in the new vacuum, they gain momentum when the bubbles expand. Therefore, the leading order contribution from the pressure onto the expanding bubble wall has the opposite sign compared to standard transitions, where the {nucleation} is towards a minimum with lesser relativistic states \cite{Buen-Abad:2023hex}. It has recently been realized \cite{Barni:2024lkj,Barni:2025mud} that this aspect is connected with more generic properties of the hydrodynamics of the plasma surrounding the expanding bubbles, which is treated as a perfect fluid affecting the dynamics of the transition and production of gravitational waves \cite{Espinosa:2010hh,Caprini:2019egz, Hindmarsh:2015qta,Hindmarsh:2017gnf,Giese:2020rtr,Giese:2020znk,Wang:2021dwl,Ajmi:2022nmq,Tenkanen:2022tly,Wang:2022lyd,Wang:2023jto}. 

Indeed, in the canonical, or direct, scenario the system enters the new phase while bubbles of the new vacuum nucleate, expand, and push the surrounding plasma outward, releasing the latent heat in an exothermic process.  

Here, instead the dynamics is as in \cite{Barni:2024lkj,Barni:2025mud}, where
the bubbles still nucleate and expand much as in the standard case, but the plasma is flowing inward. In these transitions, the latent heat is negative, and the whole process is endothermic. These transitions are called inverse.

\begin{figure}[t!]
    \centering
    \includegraphics[width=1\linewidth]{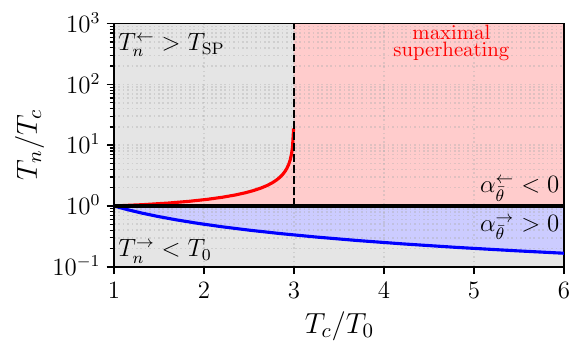}
    \caption{\it 
Allowed region for the nucleation temperature, $T_0<T_n<T_{\rm SP}$.  In the upper branch ($T_n^{\leftarrow}>T_c$), heating inverse transitions with negative latent heat ($\alpha_{\bar\theta}^{\leftarrow}<0$), while in the lower branch ($T_n^{\rightarrow}<T_c$), direct cooling transitions exhibit positive latent heat ($\alpha_{\bar\theta}^{\rightarrow}>0$). The solid lines indicate the spinodal temperature (red) and the second-order PT temperature (blue) and can be interpreted as the maximal allowed superheating (supercooling).
}
    \label{fig:inverse}
\end{figure}

The hallmark of an inverse first‐order transition is that the latent heat is negative, which in the frame of the centre of the bubble corresponds to an inward fluid velocity into the newly nucleated region. A simple diagnostic of this behaviour is the generalised pseudo‐trace of the energy–momentum tensor \cite{Barni:2025mud},
\begin{equation}
\alpha_\vartheta \equiv{D\vartheta\over 3w_+}=\frac{1}{3\,w_+}\left[D e -\frac{\delta e}{\delta p}D p\right],
\end{equation}
where $w=p+e$ is the enthalpy density and $p,e$ the pressure and energy density. The jumps are defined by $D f \equiv f_+(T_+) - f_-(T_+)$, and $\delta f \equiv f_-(T_+) - f_-(T_-)$,where $\pm$ refers to old/new phase. It is possible to show that 
\begin{equation}
\alpha_\vartheta\gtrless0 \quad\Longleftrightarrow\quad \text{direct/inverse PT} \ ,
\end{equation}
and this condition remains valid regardless of the specific equation of state or of the level of approximation. The quantity $\alpha_\vartheta$ is nothing but the thermodynamic latent heat, $L=D\vartheta$, computed away from the critical temperature, normalised with the enthalpy of the old phase.

We avoid a full hydrodynamic treatment in favour of the pseudo‐trace criterion, valid when temperature jumps across the wall are small.  Recalling that $\bar\theta = e -p/c_{s,-}^2$ and {$\alpha_{\bar\theta} =D\bar\theta/3w_+$} (see \cite{Giese:2020rtr}), one finds that the sign of \(\alpha_{\bar\theta}\) uniquely identifies the transition type.  

Figure \ref{fig:inverse} displays the allowed region in nucleation temperature.  When \(T_n^{\leftarrow}>T_c\), these are heating PTs, the latent heat is negative (\(\alpha_{\bar\theta}^{\leftarrow}<0\)), corresponding to inverse transitions—whether the barrier eventually vanishes at a finite spinodal (\(T_c/T_0<3\)) or persists indefinitely (\(T_c/T_0>3\)).  Conversely, for \(T_n^{\rightarrow}<T_c\) the cooling transitions feature \(\alpha_{\bar\theta}^{\rightarrow}>0\) and exhibit the usual direct behaviour. In these last estimates, we considered \(a_{0,2,3}\gg a_4\) so that \(c_s^2\simeq1/3\), appropriate for theories with many relativistic degrees of freedom (cf.\ Sec.~\ref{sec:scaleinv}).  The solid red (blue) line indicating the spinodal ($T=T_0$) temperature can be regarded as the maximal superheating (supercooling) we can achieve.

In conclusion, we find that all the phase transitions occurring while heating in our study are of the inverse type, regardless of the parameter values considered. On the other hand, the transitions that take place while cooling remain direct in the limit of a large number of degrees of freedom.

\section{Gravitational waves from superheating}\label{sec:applications}
We have identified the conditions under which a scalar $\phi$ can find itself in a metastable minimum at high-temperature. We have not committed to a particular realisation, but we believe our findings can be applied to many scenarios, for example to the Higgs during a modified electroweak phase transition \cite{Grojean:2004xa,Delaunay:2007wb,Ellis:2018mja,Ellis:2019oqb,Bruggisser:2018mus,  Espinosa:2007qk, Beniwal:2017eik,Barger:2007im,Espinosa:2011ax,Kozaczuk:2019pet, Kurup:2017dzf,Azatov:2022tii}. Here we highlight the main phenomenological consequence of gravitational wave production, leaving the technical details to future work.

The physics of the gravitational waves produced at the superheating phase transitions depends strongly on the details of the reheating sector of section \ref{sec:reheating}. Here we compute only the contribution from the vacuum, in order to get estimates for the amplitude and the frequency at the peak. We follow \cite{Athron:2023xlk} for the derivation.

Here we notice that the temperature rise happens when the universe finds itself in a phase of matter dominance - driven by the reheating sector - and just when $H(t)\approx \Gamma_R$ the universe is in radiation at global temperature $T_R$.\footnote{This is the usual reheating temperature of the universe.} Clearly $T_{\rm max}(t_{\rm max})\ll T_R$, and the system is also potentially subject to dilution if $t_{\rm max}$ is an epoch well before the start of radiation dominance.

In section \ref{sec:nucleation} we have identified several possible phase transitions, and we can say that they happen during the following phases
\begin{equation}
    \left\{ \begin{array}{cc}
       (\leftarrow, h)  & \text{matter dominance} \\
        (\leftarrow, c)  & \text{matter/radiation dominance} \\
         (\rightarrow, c)  & \text{mostly likely radiation  dominance} \\
    \end{array}
    \right.
\end{equation}
The inverse transition while heating certainly happens during matter dominance, since the temperature raise is present only for time scales much shorter than the decay rate $\Gamma_R^{-1}$. The inverse transition while cooling (after the system drops below $T_{\rm max}$ adiabatically), could happen during either matter or radiation dominance, depending on whether the nucleation happens before or after $T_R$.
The dilution factor $\eta$ for the inverse (while heating) is the fourth power of the ratio of the scale factor at nucleation and at $T_R$, leading to the estimate
\begin{equation}
    \eta^{\leftarrow} \approx \mathrm{min}[1, (H_R/H_*)^{8/3}]\,.
\end{equation}
The dilution factor determines the amplitude of the gravitational wave produced by the inverse phase transition as 
\begin{equation}
\begin{split}
\Omega_{\rm gw}^{\leftarrow}  &= \Omega_\gamma \frac{\rho_{\rm gw}(t^{\leftarrow}_n) \eta}{\rho_R(T_R)} \frac{g_*(T_R)}{g_*(T_\gamma)}\left(\frac{g_*^s(T_\gamma)}{g_*^s(T_R)}\right)^{\frac{4}{3}}\\
&\approx 10^{-6}\,  c_V v_w^3   \left(\frac{200}{g_*(T_R)}\right)^{\frac13} \times \left(\frac{H}{\beta}\right)^2 \frac{V_{\rm min}^2}{\rho_R(T_R)^2} \,\times \eta^{\frac 74} \,, 
\end{split}
\end{equation}
where we see that $\Omega_{\rm gw}\propto T_n^8/T_R^8$. Here $g_*$ is the relativistic number of degrees of freedom, $\Omega_\gamma$ and $T_\gamma$ are the abundance and temperature of the photons today, while $v_w$ is the velocity of the bubble wall. The direct $\rightarrow$ transition is most likely to happen during radiation dominance and not subject (or not so much) to dilution, and it is given by the above amplitude with $\eta=1$.

\begin{figure}
    \centering
    \includegraphics[width=1\linewidth]{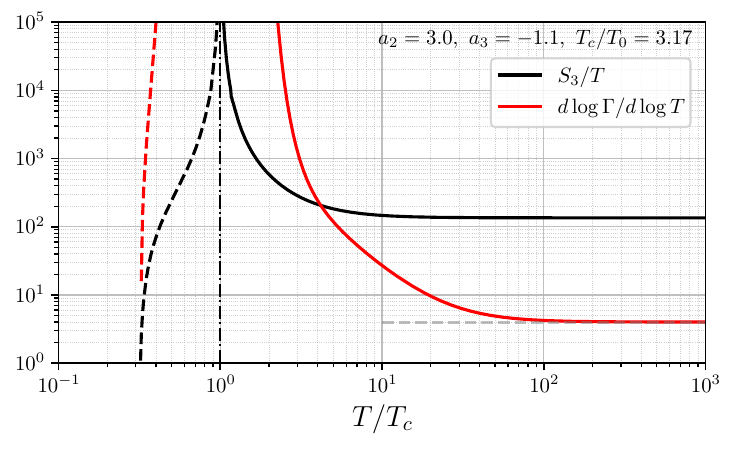}
    \caption{\it 
   Comparison between the ``intrinsic 
\(\beta/H = d\log\Gamma/d\log T\)" and the corresponding Euclidean bounce action. Solid curves denote {nucleation} toward the symmetric vacuum ($\leftarrow$), while dashed curves denote {nucleation} toward the broken vacuum ($\rightarrow$). The grey dashed horizontal line at $d\log \Gamma/d\log T=4$ indicates the minimal intrinsic value attained in the scale‐invariant regime.%
 }
    \label{fig:beta}
\end{figure}

The peak frequency of the gravitational wave spectrum is related to the value of $\beta/H$, and it is again sensitive to dilution for the inverse transitions. In general, we write it as
\be
f_{\rm peak}\sim 10\, \mathrm{Hz} \frac{\beta}{H}  \left(\frac{T_R}{10^8\, \mathrm{GeV}}\right)\times\eta^{\frac{1}{4}}\,.
\ee
In our model, the size of $\beta/H$ at the nucleation can be estimated as follows, recalling that $\beta\equiv (d\Gamma/dt)/\Gamma$. In our context, the time evolution is not monotonic in terms of the evolution of the temperature of our sector. Therefore we expect new features in the value of $[\beta/H]_n$, depends on the temperature evolution as
\begin{equation}
  \frac{\beta}{H}\bigg|_n=\left[\frac{T}{\Gamma}\frac{d\Gamma}{dT}\right]_n \times \left(\frac{\dot{T}/T}{H }\right)_n\ .
\end{equation}
The last term in parentheses has different values for the different types of transitions. For direct transition while cooling, it equates to $-1$, while a different factor arises for the inverse transitions $(\leftarrow,h)$ and $(\leftarrow,c)$. In particular, we find
\be
\frac{\beta}{H}\bigg|_n^{\leftarrow,h}\approx O(10)\times \frac{1}{t_n H_*^{-1}}\ ,
\ee
where $O(10)$ comes from the fact that the transition happens when $S_3/T$ approaches its asymptotic almost constant value. We see that the fact that the temperature of the sector is rising on time scales shorter than Hubble and $\Gamma_R$, makes $\beta/H$ effectively bigger by a factor $H_*/\Gamma_R$. To support our estimate we compute $d\log \Gamma/d\log T$ in figure \ref{fig:beta}. Here $d\log \Gamma/d\log T$ can be thought of as an intrinsic value for $\beta/H$ before the time dependence of the temperature is taken into account.

\section{Conclusions}\label{sec:conclusions}
In this paper, we have introduced the concept of super‐heating in cosmological phase transitions.  Our benchmarks are sectors with a large number of bosons, with classical scale invariance, and we have identified the broad conditions that allow a metastable vacuum to persist up to arbitrarily high temperatures. 

During super-heating (or simply heating), these sectors have an effective potential that can give rise to an inverse first‐order phase transition upon heating ($\leftarrow$), followed inevitably by the corresponding direct transition during cooling ($\rightarrow$). The former proceeds towards a vacuum with massless states, the latter towards a vacuum with massive states.

All the phase transitions towards the origin ($\leftarrow$) are of inverse type, and their corresponding ones from the origin ($\rightarrow$) are direct. By computing the bounce action $S_3/T$, the nucleation criterion from Eq.~\eqref{eq:nucleation} shows that even if an inverse phase transition while heating does not occur, it may happen during the subsequent cooling. 

In our scenario, the inverse transition induced by heating always takes place during a matter‑dominated era due to the slow heating, while the one upon cooling depends on the reheating model. If it occurs, then the corresponding direct transition while cooling inevitably follows. As noticed in \cite{Buen-Abad:2023hex}, we find that even more generally, these pairs of phase transitions could generate a distinctive double‑peaked gravitational‑wave spectrum. The first peak, if arising in the matter dominated phase, receives additional redshift dilution, while the second peak originates from the familiar radiation‑dominated direct transition. Furthermore, by estimating the phase‑transition duration we demonstrate that a rapid temperature rise enhances $\beta/H$ compared to usual phase transitions. We have emphasised the importance of establishing thermal equilibrium well before the temperature where the tree-level instability disappears, and this is correlated with the prediction for gravitational waves.

Super-heating can also be present in more realistic scenarios with both fermionic and bosonic degrees of freedom, provided that their interactions do not generate a spinodal temperature. Then, we believe that more generic scenarios can be constructed with a sizeable amount of super-heating, and what we have presented here has the minimal set of ingredients.

Further directions are also possible.  First, the ability to maintain symmetry non‐restoration up to arbitrarily high temperatures suggests other realisations of electroweak baryogenesis. Second, a dedicated numerical study of gravitational‐wave production—accounting for the two successive transitions and the modified hydrodynamics—is essential to refine the predicted double‐peaked spectrum and assess detectability in upcoming observatories.  Also, from a theory perspective, it would be desirable to identify more general scenarios. This could be pursued both by relaxing the large‑\(N\) or by fully exploiting the symmetry argument, for instance by just discussing the properties of CFT operators in the background of the field $\phi$ at finite temperature. 

\subsubsection*{Acknowledgements}
{\small 
The work of AT is supported in part by the Italian Ministry of University and Research (MUR) through the PRIN 2022 project n. 20228WHTYC (CUP:I53C24002320006). GB is supported by the grant CNS2023-145069 funded by MICIU/AEI/10.13039/501100011033
and by the European Union NextGenerationEU/PRTR. GB also acknowledges the support of the Spanish Agencia
Estatal de Investigacion through the grant “IFT Centro
de Excelencia Severo Ochoa CEX2020-001007-S”. }

\bibliographystyle{jhep}
\small
\bibliography{biblio}
\end{document}